\documentclass[aps,prl,twocolumn,showpacs,groupedaddress]{revtex4}
\usepackage{graphics}
\begin{document}

\title{Autler-Townes splitting in two-color photoassociation of $\bf^6$Li}

\author{U. Schl\"oder, T. Deuschle, C. Silber, and C. Zimmermann}

\affiliation{Physikalisches Institut, Eberhard Karls Universit\"at, Auf der Morgenstelle 14, 72076 T\"ubingen, Germany}

\date{\today}

\begin{abstract}
We report on high-resolution two-color photoassociation spectroscopy in the triplet system of
magneto-optically trapped $^6$Li. Photoassociation is induced from the ground-state asymptote to the $v\!=\!59$ level of the excited
1$^3{\rm\Sigma}^+_g$ state. This level is coupled to the $v\!=\!9$ level of the
triplet ground state $a^3{\rm\Sigma}^+_u$ by the light field of a
Raman laser. The absolute transition frequencies are measured with an iodine
frequency standard and the binding energy of the ground state is determined to
$24.391\pm0.020$\,GHz. Strong coupling of the bound molecular states has been observed as
Autler-Townes splitting in the photoassociation signal for different detunings
of the Raman laser. From the splitting we determine the spontaneous bound-bound
decay rate to  $4.3\times10^5$\,s$^{-1}$ and estimate the molecule formation rate. The observed
lineshapes are in good agreement with the theoretical model.
\end{abstract}

\pacs{32.80.Pj, 33.80.Ps, 34.20.Cf, 42.50.Hz}

\maketitle

\section{}
Coherent coupling schemes offer
intriguing novel possibilities for the production of cold molecules. In such
schemes, the continuum state of a free atom pair is coupled to a rovibrational level of the molecular
ground state. Magnetic coupling in the vicinity of
a Feshbach resonance has led to the observation of atom-molecule coherence in a
$^{85}$Rb BEC~\cite{don} and to the selective production of weakly bound Cs$_2$ dimers~\cite{chi}. Optical coupling in stimulated
Raman transitions has allowed for the creation of $^{87}$Rb molecules in a
BEC~\cite{wyn} and of Cs$_2$ molecules from a magneto-optically trapped atomic cloud~\cite{lab}. In
contrast to single-color photoassociation experiments, where cold ground-state
molecules are produced incoherently by spontaneous decay of the optically excited dimer~\cite{fio,nik,gab}, these
transitions schemes have coherent character. Thereby, ground-state molecules
are produced in a specific rovibrational level. In addition,
the molecule formation rate can be strongly enhanced, as it is not limited by the small branching ratio between
the bound-bound and the bound-free transitions. This is especially important for the formation of dimers with small intrinsic molecule
formation rates, as in case of the absence of any favorable peculiarities in their
potentials~\cite{fio,gab}. While magnetic coupling schemes require a Feshbach
resonance at sufficiently low magnetic fields and allow only for the production
of cold molecules in weakly bound states, optical coupling schemes are more
universal. Moreover, in extended multi-color versions,
they are proposed for the production of molecules which are not only
\begin{figure}[b]
\resizebox{0.5\textwidth}{!}{\includegraphics{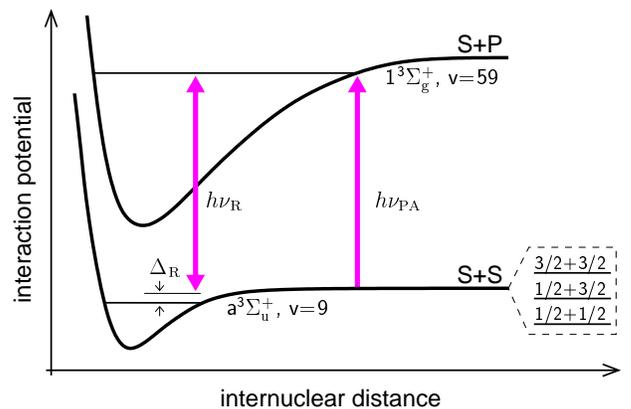}}
\caption{Coupling scheme for the stimulated Raman photoassociation in the triplet system of
the $^6$Li$_2$ molecule. Photoassociation is induced from the ground-state asymptote $1/2+3/2$ into the hyperfine state $|1111\rangle$ of the
excited level $v\!=\!59$. The Raman laser couples this excited level to the
hyperfine state $|0111\rangle$ of the ground-state vibrational level $v\!=\!9$.}
\end{figure}
translationally, but also vibrationally cold~\cite{cot}.\\
Among the cold homonuclear dimers, $^6$Li$_2$ molecules in the triplet state
are particularly interesting. Combining two $^6$Li atoms to a
$^6$Li$_2$ molecule corresponds to transfering a Fermionic system into a
Bosonic system. Such transition schemes are prerequisite for studying the effects of different
quantum statistics on the reaction dynamics, the so-called
superchemistry~\cite{hei}. If, in addition, the molecule is prepared in the
triplet state, it has a magnetic dipole
moment. Storing these cold molecules in a magnetic trap should be possible, as
demonstrated for cesium triplet dimers in a magnetic quadrupole trap recently~\cite{van}.\\
In this paper we report on two-color photoassociation in the triplet system of
$^6$Li$_2$. We use a Raman-type coupling scheme as shown in Fig.\,1. Photoassociation is induced
from the hyperfine ground-state asymptote $f\!=\!1/2+f\!=\!3/2$ into the
excited triplet 1$^3{\rm\Sigma}^+_g$ state by the
photoassociation laser with frequency $\nu_{\rm PA}$. From the excited rovibrational
level $v\!=\!59$, $N\!=\!1$, the specific hyperfine state $|1111\rangle$
($|NSIG\rangle$-notation, see e.g.~\cite{abr}) is selected. The excited level lies at a laser detuning of $-1787$\,GHz relative
to the $D1$-line, corresponding to a Condon point of about 34\,a$_{0}$. The
natural linewidth is approximately twice the atomic value $\Gamma$ and
amounts to 2$\pi\times$11.7\,MHz~\cite{cot1}. The Raman laser of the frequency $\nu_{\rm R}$
couples the excited state to the hyperfine state $|0111\rangle$ of the $v\!=\!9$,
$N\!=\!0$ level of the
triplet ground state $a^3{\rm\Sigma}^+_u$. The
ground-state level has a binding energy of approximately $-24$\,GHz and an outer turning
point of 27\,$a_{0}$. The binding energies and the hyperfine structure involved
in this scheme have been determined in photoassociation experiments of R. Hulet
and co-workers~\cite{abr,abr1}. In our experiments we focus on an absolute frequency
measurement of the bound-free and the bound-bound transition and therefrom we deduce the binding energy of the ground
state. Furthermore, we study the influence of the
detunig of the fixed Raman laser on the photoassociation
signal. Autler-Townes splitting in the two-color photoassociation signal is observed,
which arises from the strong coupling of the bound molecular states. We discuss
the lineshape in the light of the theory of Bohn and
Julienne~\cite{boh}. From the amount of
the splitting we deduce the spontaneous decay rate and estimate the molecule
formation rate. These results on the Autler-Townes splitting are complementary
to recent two-color photoassociation experiments with cesium~\cite{lab}.\\ 
For the magneto-optical trap we use the two-species apparatus as
described in~\cite{sch}, but operate it with $^6$Li
only. We apply a detuning of $-5$\,$\Gamma$ for the cooling
transition and of $-2$\,$\Gamma$ for the repumping transition. The temperature of
the atomic ensemble can be estimated to 0.5\,mK~\cite{schu}. The particle number is monitored by
absorption from a weak probe beam.  For the
photoassociation light we use a grating-stabilized diode laser with an output power of 12\,mW. The light of the Raman beam is derived from a dye laser with 500\,mW
output power. The two laser beams are
superimposed with parallel polarization and adjustable relative laser
powers. The 1/$e^2$ diameters of the laser beams at the position of the atomic
cloud amount to 550\,$\mu$m, corresponding to peak intensities of up to $I_{\rm
R}\!=\!40$\,W/cm$^2$
for the Raman laser and of up to $I_{\rm PA}\!=\!7$\,W/cm$^2$ for the photoassociation laser. The
stabilization and tuning of the diode laser is achieved by means of a
stabilization chain. The diode is locked to a Fabry-Perot etalon with 300\,MHz free spectral
range. This etalon is stabilized relative to a second diode laser, which is again
stabilized to a reference diode laser by a heterodyne technique. This reference laser is locked to an atomic lithium resonance by
radio-frequency sideband spectroscopy in a lithium cell. Absolute frequency calibration is achieved
by a Doppler-free iodine fluorescence spectroscopy with an accuracy of 10\,MHz~\cite{sil}. The scan rate is
chosen to 0.5-1\,MHz/s for the absolute frequency measurements and to 0.36\,MHz/s for the
Autler-Townes measurements. This is slow as compared to the loading time of the
\begin{figure}[t]
\resizebox{0.5\textwidth}{!}{\includegraphics{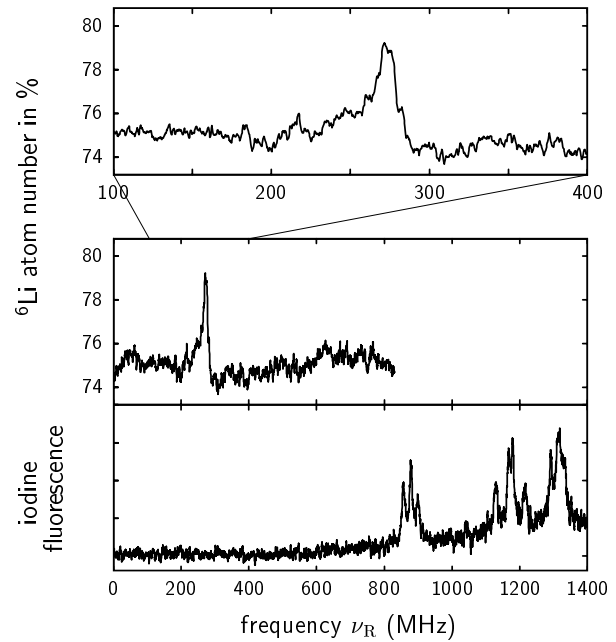}}
\caption{Two-color photoassociation spectrum and iodine reference spectrum in
the case of a fixed photoassociation laser and a scanned Raman laser. The
offset to the axis of the Raman laser frequency $\nu_{\rm R}$ is 445027\,GHz.}
\end{figure}
trap of 15\,s. Thereby a resolution of a few MHz is achieved.\\ 
For the absolute frequency measurement of the photoassociation and the
Raman transition we proceed in the following manner. First, we record the
single-color photoassociation spectrum and the iodine spectrum in the relevant
frequency range simultaneously. For this special case we use the dye laser to
induce photoassociation because of its larger tuning range. In order to keep the influence of
the light shift~\cite{boh1} small, we reduce the laser intensity to
3\,W/cm$^2$. For the transition
from the hyperfine ground-state asymptote $f\!=\!1/2+f\!=\!3/2$ into the
specific excited hyperfine state we determine the frequency to
$445002.881\pm0.015$\,GHz. A redshift of 10\,MHz due to the thermal energy of
the atoms has been included in this analysis. Then, the photoassociation laser is held fixed on
this transition, which leads to a decrease in the steady-state particle number of the trapped
atoms to 75\%. The Raman laser is scanned with an intensity of 10\,W/cm$^2$. An
increase in the particle number reveals resonance
frequency of the Raman laser with the bound-bound transition, which lies at
$445027.272\pm0.013$\,GHz (Fig.\,2). At our relatively high Raman laser
intensities, this reduction of decrease originates from the strong
coupling of the bound molecular states by the Raman laser (see below), as a
consequence of which the photoassociation laser is no longer resonant with the
free-bound transition and particle losses due to photoassociation are suppressed.
 The binding
energy of the bound ground-state level $v\!=\!9$ in the absence of hyperfine structure is equivalent to the difference between the measured transition
frequencies, because the initial and final level are shifted due to the hyperfine
structure by the same amount. We obtain a value of $24.391\pm0.020$\,GHz. 
The uncertainty is mainly due to a possible
systematic error in the iodine reference spectrum and moreover due to the error in
the determination of the line position. Our result is in
agreement with the value of $24.43\pm0.02$\,GHz obtained from a direct
\begin{figure}[t]
\resizebox{0.5\textwidth}{!}{\includegraphics{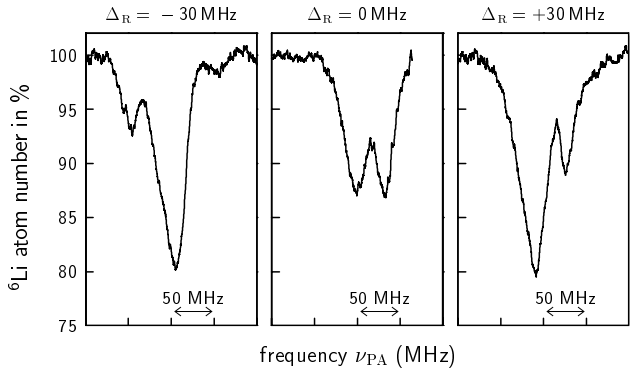}}
\caption{Autler-Townes splitting for three different detunings $\Delta_{\rm
R}$ of the Raman laser. The axis of the
photoassociation laser frequency $\nu_{\rm PA}$ shows relative frequency calibration.}
\end{figure}
measurement of the frequency difference~\cite{abr}.\\
The lineshape of the two-photon signal shows strong asymmetries with a shallow slope at the
low frequency side and a steep slope at the high frequency side. We determine
the linewidth (full width at half maximum) to 15\,MHz. Therewith, the linewidth
is smaller than the convolution of the natural linewidth of 11.7\,MHz and the thermal broadening of about 10\,MHz. The lineshape is that
of a typical Fano
profile~\cite{fan,coh}, which has been analyzed experimentally in~\cite{lis}.\\
For the Autler-Townes measurements, we first tune the Raman laser into resonance with
the bound-bound transition for calibration. Then, a fixed detuning
$\Delta_{\rm R}$ is added and the photoassociation laser is scanned. Two-color photoassociation
spectra for three different detunings $\Delta_{\rm R}$ of the Raman
laser are shown in Fig.\,3. The spectrum in the middle has been taken in presence
of a resonant Raman laser ($\Delta_{\rm R}\!=\!0$\,MHz). For the outer spectra, the
Raman laser was detuned by the same amount of 30\,MHz to the red and to
the blue respectively. For these spectra, the laser intensities were chosen to
30\,W/cm$^2$ for the Raman laser and to 3\,W/cm$^2$ for the photoassociation laser. All three resonances show a doublet structure with a splitting of several 10\,MHz. In the case of the resonant
Raman laser, the resonance is split into two peaks with equal depth. For a
detuned Raman laser, the splitting is asymmetric with a larger peak on the
high-frequency side for red detuning and with a larger peak on the
low-frequency side for blue detuning. To determine the
splitting we fit two Lorentzian
functions to the data. For the resonant Raman laser the spacing between the maxima of the resonances amounts
to 34.6\,MHz. For the detuned Raman laser the spacing is 50.1\,MHz and
40.4\,MHz respectively.\\
The origin of the splitting can be explained in the dressed-state
picture~\cite{coh}. Here, the two bound molecular states are dressed with the light
field of the Raman laser, forming a ladder of doublets that are separated by the generalized Rabi frequency $\Omega_{\rm gen}/2\pi$. Its value
depends on the detuning $\Delta_{\rm R}$ of the Raman laser and of the Rabi
frequency $\Omega$ (see below) according to $\Omega_{\rm
gen}/2\pi\!=\!\sqrt{\Delta_{\rm R}^2+(\Omega/2\pi)^2}$. By
spectroscopy with a weak laser beam from
a third level this splitting appears as an Autler-Townes doublet~\cite{aut}. In
the case considered here, this third level corresponds to the continuum state.\\ 
\begin{figure}[t]
\resizebox{0.5\textwidth}{!}{\includegraphics{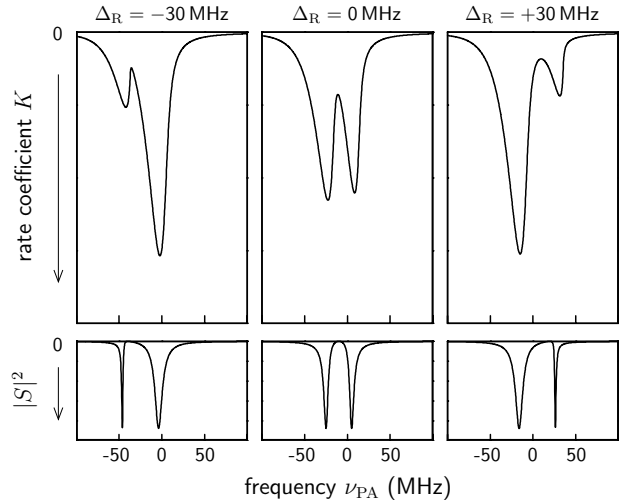}}
\caption{Scattering probability $|S|^2$ for a scattering
energy of $E/h\!=\!10\,$MHz and rate
coefficient $K$ for a temperature of 0.5\,mK for three
different detunings $\Delta_{\rm R}$ of the Raman laser. For the
Rabi frequency $\Omega\!=\!2\pi\times30$\,MHz and for the natural linewidth $\gamma\!=\!12$\,MHz were assumed. The frequency axis is relative to the transition
frequency for vanishing energy. To compare the
calculation with the experimental data, the ordinate is plotted upside-down.}
\end{figure}
A more sophisticated model, also including lineshapes, is given by Bohn and
Julienne~\cite{boh}. This description is based on scattering theory and yields a
general analytic expression for the two-color photoassociation scattering probability
$|S|^2$. For a thermal ensemble, the corresponding rate coefficient $K$ is
obtained by thermal averaging. The results of a simulation for the corresponding parameters of our experiment are shown in
Fig.\,4. In this theory, the
origin of the splitting is explained by two maxima of the scattering
probability $|S|^2$ as a function of the photoassociation laser frequency
$\nu_{\rm PA}$. The
magnitude of the splitting is independent of the scattering energy $E$ and
corresponds to the generalized Rabi frequency $\Omega_{\rm gen}$. By thermal averaging over the
scattering energies, the different resonance widths in the scattering
probability $|S|^2$ translate
into different heights of the doublet peaks of the rate coefficient $K$. However, thermal averaging does not
change the amount of the splitting substantially, which we have checked for
various parameters. The thermal averaged resonances are asymmetric. The high-frequency sides are always steeper. This is due to the convolution of the small
intrinsic linewidth of the transition with the Boltzmann distribution of the
scattering energies $E$. Thereby, contributions from low energies, corresponding
to higher frequencies, are favored.\\ 
The measured trap loss (Fig.\,3) is proportional to the calculated
photoassociation rate
coefficient $K$ (Fig.\,4), if a linear response of the steady-state particle
number to the rate coefficient $K$ is assumed. For low photoassociation laser intensities this approximation
is justified. The agreement concerning the splitting and the relative heights of
the peaks between experiment and theory is quite good. Moreover, the asymmetry
due to the thermal averaging is visible in the experimental spectra,
particularly for the spectrum with $\Delta_{\rm R}\!=\!-30$\,MHz.\\ 
For the experimental determination of the Rabi frequency $\Omega$, we use the
data taken for a resonant Raman laser. In this case the generalized Rabi frequency
$\Omega_{\rm gen}$ is identical to the  Rabi frequency $\Omega$ and
independent of uncertainties in the detuning of the Raman laser. Therefore,
the measured splitting of 34.6\,MHz  can be identified with the Rabi frequency $\Omega/2\pi$. By
using the relation for the Rabi frequency \mbox{$\Omega\!=\!\sqrt{3 c^2 I_{\rm
R}\Gamma_{\rm bb}/4\pi h\nu_{\rm R}^3}$}, we determine the
\mbox{transition rate $\Gamma_{\rm bb}$} between the bound states to
\mbox{$4.3\times10^5$\,s$^{-1}$}. This coincides well with a value of
$6.16\times10^5$\,s$^{-1}$~\cite{cot}, derived theoretically from a simplified
two-level system.\\ 
The experimental data allow an estimation of the molecular formation rate per atom and
per second. For the resonant
Raman laser, each doublet peak consists of contributions from the
molecular ground state and from the excited state to equal parts. Thereby, the molecule formation rate should
be given by half the single-color photoassociation rate. The
photoassociation rate itself can be estimated from the 87\% reduction of the steady-state
number of $^6$Li atoms in presence of the photoassociation laser (see
Fig.\,3). Assuming constant density and using the measured value for the
loading time of the trap, we use the formula given in~\cite{sch1}. We receive a
value for the photoassociation rate in the range of 0.01\,s$^{-1}$ per atom, leading to a molecule formation rate in the
same order of magnitude. This is an enhancement by a factor of 100 compared to
the molecule formation rate due to single-color photoassociation, which can be
estimated to $8.4\times10^{-5}$\,s$^{-1}$ by multiplying the photoassociation rate with the branching ratio between the bound-bound
and the free-bound transitions~\cite{cot}. For typical particle numbers in the
order of 10$^8$ atoms~\cite{sch},
the absolute molecule formation rate amounts to 10$^6$\,molecules/s. However,
in order to estimate the number of formed molecules, loss processes, such as
spontaneous Raman scattering, have to be taken into account.\\
In conclusion, we have performed high-resolution two-color photoassociation
spectroscopy in the triplet system of the $^6$Li$_2$ molecule. The binding
energy of the last bound level of the ground state has been determined in an
absolute frequency measurement. We have observed the Autler-Townes splitting in
the two-color photoassociation signal and therefrom deduced the bound-bound
transition rate and the molecule formation rate. This rate is sufficiently
high to encourage future experiments in this transition scheme. These are the
accumulation of the cold triplet dimers in a magnetic trap and the production of
vibrationally cold ground-state molecules via multi-color photoassociation schemes. For the degenerate regime, even pulsed
schemes such as STIRAP~\cite{var} are discussed.
\begin{acknowledgments}
We would like to thank H. Schnatz and B. Bodermann from the PTB
Braunschweig and E. Tiemann for valuable help with the iodine spectroscopy. This work has been partially funded by the Deutsche Forschungsgemeinschaft.
\end{acknowledgments}


\end{document}